# There is no vacuum zero-point energy in our universe for massive particles within the scope of relativistic quantum mechanics

Huai-Yu Wang
*Department of Physics, Tsinghua University, Beijing 100084, China*
wanghuaiyu@mail.tsinghua.edu.cn
794024763@qq.com phone number: +86-13717938266

**Abstract:** It was long believed that there is a zero-point energy in the form of $\hbar\omega/2$ for massive particles, which is obtained from Schrödinger equation for the harmonic oscillator model. In this paper, it is shown, by the Dirac oscillator, that there is no such a zero-point energy. It is argued that when a particle's wave function can spread in the whole space, it can be static. This does neither violate wave-particle duality nor uncertainty relationship. Dirac equation correctly describes physical reality, while Schrödinger equation does not when it is not the nonrelativistic approximation of Dirac equation with a certain model. The conclusion that there is no zero-point energy in the form of $\hbar\omega/2$ is applied to solve the famous cosmological constant problem for massive particles.

**Résumé:** On a longtemps cru qu'il existait une énergie du point zéro sous la forme de $\hbar\omega/2$ pour les particules massives, qui est obtenue à partir de l'équation de Schrödinger pour le modèle d'oscillateur harmonique. Dans cet article, il est montré, par l'oscillateur de Dirac, qu'il n'y a pas une telle énergie du point zéro. On fait valoir que lorsque la fonction d'onde d'une particule peut se propager dans tout l'espace, elle peut être statique. Cela ne viole ni la dualité onde-particule ni la relation d'incertitude. L'équation de Dirac décrit correctement la réalité physique, contrairement à l'équation de Schrödinger lorsqu'il ne s'agit pas de l'approximation non relativiste de l'équation de Dirac avec un certain modèle. La conclusion qu'il n'y a pas d'énergie du point zéro sous la forme de $\hbar\omega/2$ est appliquée pour résoudre le fameux problème de la constante cosmologique pour les particules massives.

# I. INTRODUCTION

As soon as Schrödinger had established his stationary equation for wave mechanics,[1] he solved the harmonic oscillator model,[2] and the energy spectrum showed a lowest energy in the form of $\hbar\omega/2$ which was called zero-point energy (ZPE). Since then, it has been believed that in quantum mechanics (QM), there is a ZPE for the motion of a particle. Usually, in QM textbooks, after introducing Schrödinger equation, the harmonic oscillator model and the concept of the ZPE were presented.[3-15]

The above concept of ZPE was from the oscillator model. Let us inspect the case of free particles. From Schrödinger equation, a free particle does not have the ZPE. However, for relativistic particles, the ZPE emerges after second quantization,[16] and this ZPE is believed to be owned by both spin-0 and spin-1/2 particles. The Schrödinger equation was believed describing the low-momentum motion of particles. Here the "low-momentum" is the synonym of "nonrelativistic", and I prefer using the former. The relativistic and low-momentum free particles have discrepancy in the property of the ZPE. This, therefore, is thought an inconsistence.

As a matter of fact, this inconsistence also appears in the case of harmonic oscillator models. As mentioned above, a one-dimensional Schrödinger oscillator has a ZPE in the form of $\hbar\omega/2$, but a Dirac oscillator does not.[17]

The above mentioned cases are listed in TABLE I. In this table, five cases are labeled by (1)-(5), respectively. In my opinion, Cases (1), (2) and (3) should be the same, i. e., all either have the ZPE or not; Cases (4) and (5) should be the same. The reason is as follows. Let us consider a free particle moving in space. Its momentum can be any value. In the whole momentum range, from zero to anbitrarily large momentum, a particle's motion either has the ZPE or not. One is unable to tell at which momentum the motion with the ZPE transits to that without the ZPE. The same reasoning applies to cases (4) and (5).

TABLE I. The cases where there is or there is no zero-point energy (ZPE).

|  | Schrödinger particle | Relativistic particle with spin-0 | Relativistic particle with spin-1/2 |
|---|---|---|---|
| Free particle | No. (1) | Yes. (2) (after second quantization)[16] | Yes. (3) (after second quantization)[16] |
| Particle subject to a harmonic oscillator potential | Yes. (4) |  | No. (5) Dirac oscillator[17,18] |

The inconsistence in TABLE I must be eliminated. This task inevitably involves the careful examination of the fundamental equations of QM.

The Schrödinger equation was the first QM equation. There followed two relativistic quantum mechanics equations (RQMEs), Klein-Gordon equation and Dirac equation. After the RQMEs had appeared, people realized that the Schrödinger equation could be achieved from a low-momentum approximation of the RQMEs. This reminds us the relationship between Newtonian mechanics and special relativity in classical mechanics. Historically, Newtonian mechanics was established before the special relativity. From Newtonian mechanics itself, no one was able to perceive any defect of it. Only after special relativity had emerged and its low-momentum approximation been taken, people knew that Newtonian mechanics had a limit that it stood for low-momentum motion, not for any momentum value. Similarly, from Schrödinger equation itself, no one can find any defect. Only taking low-momentum approximation from a RQME can one find what is lost in Schrödinger equation. People may say that we have already known that the Schrödinger equation is a low-momentum approximation of the RQMEs. This seems to prompt us that the relationship between the Schrödinger equation and RQMEs is clear, but it is insufficient. The RQMEs have new ingredients that special relativity in classical mechanics does not: negative kinetic energy (NKE) solutions and spin. Up to now, the author has not considered the factor of spin yet because magnetic field has not been taken into account. When one inspects the possible shortcomings of the Schrödinger equation, he has to consider the factor of the NKE solutions.

In a previous paper of the author, the fundamental equations of QM were examined carefully.[19] A remarkable discrepancy of the Schrödinger equation and RQMEs is that the latter have both positive kinetic energy (PKE) and NKE solutions while the former is not. The author had a few basic point of view. One of them was that the RQMEs were of symmetry with respect to PKE and NKE, and the PKE and NKE solutions ought to be treated on an equal footing. Specifically, if a particle with energy $E$ moves in space where the potential $V$ was piecewise constant, the PKE (NKE) solutions should be employed in regions of $E>V$ ($E<V$). Under this criterion, the famous Klein's paradox was solved perfectly, and a uniform routine was applied to both cases of particles with spin-0 and spin-1/2.[20]

However, the Schrödinger equation, as a low-momentum approximation of the RMQEs, did not reflect the symmetry with respect to the PKE and NKE. Let us retrospect how the Schrödinger equation can be derived from the RQMEs. Denote the wave function of Dirac equation by $\Psi$. Then one can make transformation $\Psi = \psi_{(+)} e^{-imc^2 t/\hbar}$ and take low-momentum approximation, so as to obtain the Schrödinger equation that the wave function $\psi_{(+)}$ obeys. Nevertheless, the kinetic energy in the Schrödinger equation is positive, so it is actually the low-momentum approximation of the PKE branch of Dirac equation and merely applies to the regions where $E>V$. The author pointed out that whether Schrödinger equation was applicable to the $E<V$ regions or not had never been verified experimentally quantitatively nor derived theoretically.[19] It was disclosed that when one took the transformation $\Psi = \psi_{(-)} e^{imc^2 t/\hbar}$ and low-momentum approximation, he would gain NKE Schrödinger equation which the wave function $\psi_{(-)}$ obeyed. The experiments of detecting NKE electrons were suggested.[19] The combination of the Schrödinger equation and NKE Schrödinger equation inherited the properties that RQMEs have. It was argued that the NKE solutions were not antiparticles, see appendices in Refs. 21 and 22. The statistical mechanics and thermodynamics of the NKE systems were elaborated[21] and related macroscopic mechanics was given.[23]

As soon as the status of the NKE was promoted to the same as that of the PKE, the field of QM was extended to the domain of the NKE, and a series work needed to be done. In Appendix D of [19], 13 points were listed which were topics to be done in the following work. The points 2, 4, 6, 7 and 9 were dealt with in Refs. [20,21,23]. The present paper intends to deal with point 1. In Ref. [24], the Virial theorem was investigated comprehensively and the symmetry of the PKE and NKE was well exhibited in this theorem.

In the author's opinion, in treating low-momentum motion, one also should distinguish the PKE ($E>V$) and NKE ($E<V$) regions where the Schrödinger equation and NKE Schrödinger equation ought to be employed, respectively. However, up to now all the low-momentum motion has been solved only by Schrödinger equation even when the particle's energy $E$ was less than $V$. The author resolved a few one-dimensional problems usually treated in QM textbooks,[25] disclosing the discrepancies arising from taking into account the NKE Schrödinger equation. The reflection coefficient of one-dimensional step potential was recalculated,[22] and this treatment was in line with that for relativistic particles.[21]

The example in Ref. 22 was a simple one since the potential was piecewise constants. For a continuous potential, the piecewise treatment of the potential will be intractable.

Let us see the case of a hydrogen atom with its nucleus motionless. The Coulomb attraction is that $V(r) = -\dfrac{a}{r}$ where $a > 0$. The characteristic is that $V_{min} = V(r \to 0) \to -\infty$ and $V_{max} = V(r \to \infty) \to 0$. The eigenvalues $E$ of any bounded state is between the $V_{min}$ and $V_{max}$, $V_{min} < E < V_{max}$. Therefore, in my opinion the states should be solved piecewise: in the regions where $E > V$ and $E < V$ Schrödinger equation and NKE Schrödinger equation should be employed, respectively. Obviously, such a solving procedure is difficult to carry out. Fortunately, this problem can be solved by Dirac equation.

Here I present another basic viewpoint: every relativistic equation can take low-momentum approximation; conversely, every low-momentum equation necessarily has its corresponding relativistic equation the low-momentum approximation of which is just the primary low-momentum equation.

The Schrödinger equation with Coulomb attraction has its corresponding RQME. Dirac equation solves the problem exactly in the whole space without need of comparing the energy and potential. The eigenvalues acquired from Dirac equation can be taken low-momentum approximation and then compared to those gained from Schrödinger equation: the latter, in the case of Coulomb attraction potential, are not precise although the errors are small. The same routine is applied to harmonic oscillator model in the present work.

In Section II, the harmonic oscillator model in QM is discussed, and the the conclusion is that there is no ZPE so that the inconsistence between cases (4) and (5) in TABLE I is removed. As for the inconsistence between cases (1)-(3), it concerns second quantization, and will be treated in a later work which will be based on the investigation in [19-25]. Section III will discuss the vacuum zero-point energy (VZPE) of our universe, which is the origination of the problem of cosmological constant. Section IV is the conclusion.

## II. ZERO-POINT ENERG OF A HARMONIC OSCILLATOR

In Schrödinger equation , when a particle with mass $m$ is subject to a harmonic potential

$$V(x) = \frac{1}{2}m\omega^2 x^2, \tag{1}$$

it is called the harmonic oscillator model. Here for simplicity we merely discuss the model in one-dimensional space. Schrödinger equation with this potential reads

$$-\frac{\hbar^2}{2m}\psi''(x) + \frac{1}{2}m\omega^2 x^2 \psi(x) = E\psi(x). \tag{2}$$

The solved eigenvalues of Eq. (2) are

$$E_n = (n + \frac{1}{2})\hbar\omega. \tag{3}$$

The corresponding normalized spatial wave functions are denoted by $\kappa_n$, where $n$ means the $n$-th excited state of the harmonic oscillator. The $\kappa_n$'s are all real, and are symmetric for even $n$ and antisymmetric for odd $n$. As $n = 0$,

$$E_0 = \frac{1}{2}\hbar\omega. \tag{4}$$

This is the ground state energy. The corresponding ground state wave function is

$$\kappa_0(x) = (\frac{m\omega}{\pi\hbar})^{1/4} \exp(-\frac{m\omega}{2\hbar}x^2). \tag{5}$$

The ground state energy (4) was stressed: "we call attention to the fact that it is not zero."[26] This so-called ZPE was thought that it was due to uncertainty relation.[3-15,27,28] A typical analysis was as follows: "To make the energy small we have to make kinetic energy small. Now in order to minimize the kinetic energy (and hence the momentum) a certain extension in space is required, but at the minimum of the potential the position would be fixed at $R_0$. Therefore, a certain probability for vibration is needed which in turn means a certain increase of energy."[3] This analysis was based on that it was believed that at the ground state, momentum could not be zero.[3,4] The essential reason was the uncertainty relation, also called uncertainty principle, which was often written as[3-10,28-34]

$$\Delta x \Delta p \geq \hbar/2. \tag{6}$$

The above conclusion and discussion are from Schrödinger equation which describes the low-momentum motion of particles. As we have mentioned above, one should be cautious for the regions where energy $E$ is less than potential $V$.[19,22] Let us inspect the potential (1). In the author's opinion, in the region $x^2 < 2E/m\omega^2$, Schrödinger equation should be adopted, while in the regions $x^2 > 2E/m\omega^2$, NKE Schrödinger equation should be adopted. Such piecewise solving is apparently difficult to implement. A way to gain exact solutions is to resort to RQME. As has been mentioned in introduction, the Schrödinger equation with the harmonic potential has necessarily its corresponding RQME, the low-momentum approximation of which should be exactly the same as Eq. (2).

Lucky enough, relativistic oscillator was proposed already, called Dirac oscillator. As early as 1978, the relativistic oscillator was concerned.[35] In 1989, Moshinskyt and Szczepaniak[18] proposed the model of three-dimensional Dirac oscillator, and the name was given by them. Later, some ones researched this model,[36-39] and the investigation was introduced in detail.[40] The Dirac oscillators of two-dimensional[41] and one-dimensional[17] versions were also proposed. Here for simplicity we discuss one-dimensional Dirac oscillator. The Hamiltonian is

$$H = c\sigma_1(-i\hbar \frac{d}{dx} + A(x)) + mc^2\sigma_3, \tag{7a}$$

where

$$A(x) = -im\omega x \sigma_3. \tag{7b}$$

Here

$$\sigma_1 = \begin{pmatrix} 0 & 1 \\ 1 & 0 \end{pmatrix}, \sigma_3 = \begin{pmatrix} 1 & 0 \\ 0 & -1 \end{pmatrix}. \tag{8}$$

A remarkable feature in Dirac oscillator model is that the $A(x)$ is a linear vector potential. The wave function is written in the form of $\psi = \begin{pmatrix} \varphi \\ \chi \end{pmatrix}$ and the eigenenergy is denoted as $E$. It is easy to get the equations the two components satisfy.

$$-\frac{\hbar^2}{2m}\varphi''(x) + \frac{1}{2}m\omega^2 x^2 \varphi(x) - \frac{1}{2}\hbar\omega\varphi(x) = \frac{E^2 - m^2c^4}{2mc^2}\varphi(x). \tag{9a}$$

$$-\frac{\hbar^2}{2m}\chi''(x) + \frac{1}{2}m\omega^2 x^2 \chi(x) + \frac{1}{2}\hbar\omega\chi(x) = \frac{E^2 - m^2c^4}{2mc^2}\chi(x). \tag{9b}$$

In each component equation, there is a scalar harmonic oscillator potential, which is actually from the vector potential (7b) in Dirac equation. Comparing (9a) to (2), we immediately have

$$\frac{E^2 - m^2 c^4}{2mc^2} + \frac{1}{2}\hbar\omega = (n + \frac{1}{2})\hbar\omega. \tag{10}$$

The positive eigenenergies are determined by

$$E_n = \sqrt{m^2 c^4 + 2nmc^2 \hbar\omega}, n = 0, 1, 2, \cdots. \tag{11}$$

The components of the spinor wave function are easily written:

$$\psi_n = \begin{pmatrix} a_1 \kappa_n(x) \\ a_2 \kappa_{n-1}(x) \end{pmatrix}, \tag{12}$$

where both $\kappa_n(x)$ and $\kappa_{n-1}(x)$ are solutions of Eq. (2). Obviously, one of them is symmetric and the other antisymmetric in space. The coefficients in (12) can be calculated as

$$a_1^2 = \frac{1}{2}(1 + mc^2/E), a_2^2 = \frac{1}{2}(1 - mc^2/E). \tag{13}$$

The $a_1$ is a larger component and $a_2$ is a smaller one. Now, let us take low-momentum approximation for eigenvalues (11).

$$E_n = mc^2 + n\hbar\omega, n = 0, 1, 2, \cdots. \tag{14}$$

Alternatively, we can make low-momentum approximation in Eq. (9a) by letting $E = mc^2 + E'$ and neglecting the $E'^2$ term. Then, Eq. (9a) becomes

$$-\frac{\hbar^2}{2m}\varphi''(x) + \frac{1}{2}m\omega^2 x^2 \varphi(x) - \frac{1}{2}\hbar\omega = E'\varphi(x). \tag{15}$$

Its eigenenergies are (14), just the low-momentum approximation of Eq. (11) minus static energy.

$$E'_n = n\hbar\omega, n = 0, 1, 2, \cdots. \tag{16}$$

In Eq. (11), the lowest quantum number can be $n=0$. The ground state energy is then

$$E_0 = mc^2. \tag{17}$$

Correspondingly, the ground state wave function is

$$\psi_0 = \begin{pmatrix} \kappa_0(x) \\ 0 \end{pmatrix}, \tag{18}$$

with $\kappa_0(x)$ being Eq. (5).

It is seen that the ground state energy of Dirac oscillator is different from that of

Schrödinger oscillator, although the corresponding ground state wave functions are the same. From Eq. (2) itself, one is unable to find whether there is any defect in this equation. Dirac equation is valid for all momenta, so that Eqs. (11), (14), (17) and (18) are all correct, while Eq. (4) is not.

We now discuss the discrepancy of Schrödinger equation (2) and the relativistic equation (9a) for the larger component $\varphi$. A constant term in the left hand side of (15) disappears in the left hand side of (2). This caused an extra term in energy spectrum as in (3). At the very beginning when the Schrödinger oscillator model was proposed, no one knew this shift of potential since Dirac oscillator had not been raised yet.

Now let us see if the uncertainty relation

$$\sqrt{\overline{(x-\bar{x})^2}\overline{(p_x-\bar{p}_x)^2}} \geq \frac{\hbar}{2} \tag{19}$$

is violated in the ground state when there is no zero-point energy in the energy spectrum (17). The expression of $\kappa_0$ in (18) is just (5), from which it is easily calculated that

$$\sqrt{\overline{(x-\bar{x})^2}\overline{(p_x-\bar{p}_x)^2}} = \frac{\hbar}{2}. \tag{20}$$

It is seen that the appearance of the so-called zero-point energy in (3) was not due to uncertainty relation, but was arisen from the unknowingly shifting of a term with a factor $\hbar\omega/2$ from the left hand side to right hand side.

The ground state energy of a Dirac oscillator (17) is just a static energy, which means that the particle is static. Usually it is believed that a particle cannot be static due to wave-particle duality. Unexpectedly, the Dirac oscillator, as well as its low-momentum approximation, can be static, which does not contradict wave-particle duality. The particle is still a "wave" since its ground state wave function (5) spreads in the whole space, which is in agreement with Born's statistical explanation of wave functions. When the rest energy (17) is dropped, the ground state energy, according to (16), is zero. We have solved stationary equations above, and the whole wave function should contain a factor associated with its energy $\mathrm{e}^{-\mathrm{i}E'_n t/\hbar}$. For the ground state, $\mathrm{e}^{-\mathrm{i}E'_0 t/\hbar} = 1$ since $E'_0 = 0$, see Eq. (16), which means that even the phase remain fixed. It is really a static state.

That some ones thought that the momentum of the ground state of a harmonic oscillator could not be zero[3,4] was based on inequality (6). This is actually the simplified form of the uncertainty relation Eq. (19). This simplified form may cause misunderstanding of the uncertainty relation. By Eq. (6) one may conclude that in the ground state of a harmonic oscillator the momentum could not be zero, but by Eq. (19) there is no way to reach such a conclusion. As shown by Eq. (20), the static wave function (5) follows the uncertainty relation (19).

There are indeed cases where the ground state energies are not zero. For example, in a one-dimensional finitely deep square potential with width $a$, the corresponding wave function is $\psi_1(x) = \sqrt{\frac{2}{a}}\cos(\frac{\pi}{a}x)$ and the ground state energy is $E_1 = \frac{\pi^2\hbar^2}{2ma^2}$.

This energy can be called zero-point energy. The $E_1$ is not zero because now the particle is confined in a finite space.

By the way, we briefly mention three- and two-dimensional Dirac oscillators. Three-dimensional Dirac oscillator is much more complex because in three-dimensional space there are both spin and orbital angular momenta. Details of this model have been given in literature.[17,18,36-38] Here we only concern its ground state. The Hamiltonian is

$$H = c\boldsymbol{\alpha} \cdot (\boldsymbol{p} - \mathrm{i}m\omega \boldsymbol{r}\beta) + mc^2\beta. \tag{21}$$

It is easy to get the equations the two components satisfy.

$$[\frac{p^2}{2m} + \frac{1}{2}m\omega^2 r^2 - \frac{2\omega}{\hbar}\boldsymbol{L}\cdot\boldsymbol{S} - \frac{3}{2}\hbar\omega]\varphi = \frac{E^2 - m^2 c^4}{2mc^2}\varphi. \tag{22}$$

$$[\frac{p^2}{2m} + \frac{1}{2}m\omega^2 r^2 - \frac{2\omega}{\hbar}\boldsymbol{L}\cdot\boldsymbol{S} + \frac{3}{2}\hbar\omega]\chi = \frac{E^2 - m^2 c^4}{2mc^2}\chi. \tag{23}$$

The positive eigenenergies are determined by

$$E_n = \sqrt{m^2 c^4 + 2mc^2\hbar\omega[n + 1 + (-1)^{l+j+1/2}(j+\frac{1}{2})]}, \tag{24}$$

where $l$ is the orbital quantum number, $j=l+1/2$ and $n = 0, 1, 2, \cdots$. For the ground state, we have $l = 0$ and $n = 0$, such that

$$E_0 = mc^2. \tag{25}$$

There is no zero-point energy. The ground state function is

$$\psi_0 = \begin{pmatrix} \kappa_0(r) \\ 0 \end{pmatrix}, \kappa_0(r) = (\frac{m\omega}{\pi\hbar})^{3/4} \exp(-\frac{m\omega}{2\hbar}r^2). \tag{26}$$

Two-dimensional Dirac oscillator[42-44] can be discussed mimicking the above. In one- and two-dimensional spaces, there is no concept of spin, but in three-dimensional space, there is, so that the spin-orbital coupling appears. This coupling is reflected in Eq. (24), but does not affect the conclusion that the ground state energy is (25).

Here we stress that we are discussing the harmonic oscillator model of a massive particle. Electromagnetic fields can be quantized and composed by massless particles. We do not touch massless particles here.

# III. THE SO-CALLED VACUUM ZERO-POINT ENERGY FOR MASSIVE PARTICLES IN UNIVERSE

It is believed that there is a vacuum zero-point energy (VZPE) in our universe, which is also called cosmological constant because it may be contained in Einstein's field equation. Actually, one should distinguish VZPE as two parts: one from massive particles and the other from radiation. Here we merely discuss the VZPE for massive particles. Theoretically, the VZPE for a kind of massive particles can be estimated[45-47] by

$$\frac{1}{2}\hbar \sum_i \omega_i = \frac{\hbar c}{4\pi^2} \int_0^{k_{max}} dk k^2 \sqrt{k^2 + m^2/\hbar^2} \ . \tag{27}$$

The left hand side is the sum over all ground state energies of the oscillators.[45,46,48]

However, the value estimated in such a way was about $10^{120}$ times of that by astronomical observation.[47,49-53] This caused the famous cosmological constant problem.[45-48,2,52] Although scenarios of the elimination of cosmological constant problem were suggested,[48,54-56] this problem remains intractable.[57]

The left hand side of (27) was apparently from Eq. (4). Since we have argued, by resorting to Dirac oscillator, that there was no such a zero-point energy, the left hand side of Eq. (27) does not exist. Please note that here we mean in the whole space.

It is stressed again that we are discussing the problem of zero-point energy for massive particles, as a mass $m$ appears in Eq. (27). As for the zero-point energy of radiation, we will investigate it elsewhere.

One may think that in quantum field theory (QFT), there can be VZPE in the form of $\hbar\omega/2$. To consider this topic, let us organize the ideas from QM to QFT. Quantum mechanics emerged first, from which there were conclusions. Then, second quantization of Schrödinger equation and RQMEs was implemented, and later QFT was developed. In developing QFT, people took notice that the conclusions drawn from QFT should not contradict those from QM. Since in QM, it was thought from the harmonic oscillator model that there could be zero-point energy in the form of $\hbar\omega/2$, this result was inherited in QFT. Now that we have argued above that from RQME, there should not be such a zero-point energy, it is necessary to pursue how the VZPE emerge in QFT. That will be our later work. Before examining QFT's details, one needs to have a deeper understanding of the NKE solutions of Dirac equation and have more exploration of their applications. That is what the author has already done,[19-25] and the work continues.

According to the author's opinion, a RQME is a self-consistent theory. Therefore, any problem should be figured out within the RQMEs in which it is yielded. One example was that Klein's paradox was generated in RQMEs so that it could be solved within this field,[20] with no need of resorting to knowledge of other fields.

In present case, that there is no such a zero-point energy is determined within the scope of relativistic quantum mechanics, and there is no need to touch QFT.

## IV. CONCLUSION

We have clarified the concept of zero-point energy in quantum mechanics. Solving Dirac oscillator and taking its low-momentum approximation make us know that there is no zero-point energy for a massive particle subject to a harmonic potential. In quantum mechanics, if a particle is confined in a finite room, it must have a zero-point energy, while when its wave function spreads in the whole space, its ground state can be static, which does neither violate particle-wave duality nor uncertainty relation. In the famous cosmological constant problem, the vacuum zero-point energy was estimated based on the believed zero-point of energy of the harmonic model for massive particles. According to the argument in this paper, it is concluded that in the whole space there is no vacuum zero-point energy in the form of $\hbar\omega/2$ for massive particles. Dirac equation correctly describe physical reality.

## ACKNOWLEDGMENTS

This work was supported by the National Key Research and Development Program of China (2018YFB0704304-3).